\documentclass{ws-ijmpa}
\usepackage[super,compress]{cite}
\usepackage{graphicx,subfigure,slashed}

\def\sumint{\hbox{$\sum$}\!\!\!\!\!\!\!\int}

\def\nn{\nonumber\\}

\def\lb{\left(}
\def\rb{\right)}
\def\hmu{\hat\mu}
\def\L{\ln\frac{\hat\Lambda}{2}}
\def\Lg{\ln\frac{\hat\Lambda_g}{2}}

\def\be{\begin{eqnarray}}
\def\ee{\end{eqnarray}}
\def\del{\partial}
\def\[{\left[}
\def\]{\right]}

\def\del{\partial}

\def\p{{\boldsymbol p}}

\def\k{{\boldsymbol k}}
\def\kh{\hat{\boldsymbol{k}}}

\def\d{\text{d}}

\def\sumint{\hbox{$\sum$}\!\!\!\!\!\!\!\,{\int}}

\newcommand{\smallG}{{\scriptscriptstyle{G}}}

\begin{document}
\markboth{Nan Su}
{Recent progress in hard-thermal-loop QCD thermodynamics and collective excitations}

%
\catchline{}{}{}{}{}
%

\title{Recent progress in hard-thermal-loop QCD thermodynamics and collective excitations}

\author{NAN SU}

\address{Fakult\"at f\"ur Physik, Universit\"at Bielefeld,
\\
Universit\"atsstra{\ss}e 25, 33615 Bielefeld, Germany
\\
nansu@physik.uni-bielefeld.de}


\maketitle

\begin{history}
\received{Day Month Year}
\revised{Day Month Year}
\end{history}

\begin{abstract}
I review recent developments in QCD thermodynamics and collective excitations from the hard-thermal-loop effective theory. I begin by motivating the discussion with open questions from heavy-ion collisions. I then discuss a finite-temperature and -density calculation of QCD thermodynamics at NNLO from the hard-thermal-loop perturbation theory. Finally I discuss a recent exploration of generalizing the hard-thermal-loop framework to the (chromo)magnetic scale $g^2T$, from which a novel massless mode is uncovered.

\end{abstract}

\ccode{PACS numbers: 11.10.Wx, 12.38.Aw, 12.38.Cy, 12.38.Mh, 25.75.Nq}

\tableofcontents

\newpage

\section{Introduction}\label{sec:intro}

In this brief review, I summarize progress in recent years in the thermodynamics and collective excitations of quantum chromodynamics (QCD) based on the hard-thermal-loop (HTL) effective theory. This is a very active research area, especially due to the pending challenges from the ultra-relativistic heavy-ion collision experiments at the Relativistic Heavy Ion Collider (RHIC) at Brookhaven National Laboratory, the Large Hadron Collider (LHC) at the European Organization for Nuclear Research, as well as the forthcoming experiments and the Facility for Antiproton and Ion Research (FAIR) at the GSI Helmholtz Centre for Heavy Ion Research.

Asymptotic freedom~\cite{Gross:1973id,Politzer:1973fx} predicts that at sufficiently high temperature ($T$) and density/chemical potential ($\mu$) hadronic matter would undergo a phase transition (or a crossover) to a novel deconfined phase of quark-gluon plasma (QGP). The launch of heavy-ion experiments at the RHIC and most recently at the LHC opened an unprecedented era for the study of matter under extreme conditions: the initial temperature at the RHIC is expected to reach up to 2\,$T_c$, while that of the LHC up to 4\,$T_c$, where $T_c \sim 160$\,MeV is the pseudo-critical temperature for the QCD deconfinement transition. Strikingly, the matter created in the collisions behaves like a nearly ideal liquid characterized by a small shear viscosity, which has spoiled the naive expectation from asymptotic freedom for a weakly coupled QGP (see Ref.~\citen{Schafer:2009dj} for a review). The QCD running coupling expected in the experimental energies is on the order of unity $g\sim{\cal O}(1)$, which is some \emph{intermediate} value neither infinitesimally small nor infinitely large. There have been insights gained from the strong-coupling formalism based on the anti-de Sitter/conformal field theory (AdS/CFT) correspondence (see Ref.~\citen{CasalderreySolana:2011us} for a review), however in order to have a direct access to the physical mechanisms of the QGP thermodynamics and realtime dynamics and considering the fact that finite density and realtime dynamics are yet to be settled challenges in lattice QCD (see Refs.~\citen{Philipsen:2012nu,Meyer:2011gj} for reviews), continuum QCD methods such as resummed perturbation theory are indispensable.

Collective excitations in the QGP introduce daunting challenges in practice. In addition to the intrinsic energy scale $T$, collective behaviors of the QGP generate two thermal scales, namely the (chromo)electric scale $gT$ and the (chromo)magnetic scale $g^2T$. The electric scale is dominating at high $T$ where $g$ is small and it is well described by the hard-thermal-loop effective theory~\cite{Frenkel:1989br,Braaten:1989mz,Braaten:1990az,Taylor:1990ia,Frenkel:1991ts,Braaten:1991gm} (see also Refs.~\citen{Blaizot:2001nr,Blaizot:2003tw,Kraemmer:2003gd,Andersen:2004fp,Mrowczynski:2005ki,Arnold:2007pg,Su:2012iy} for reviews on various aspects of the developments). The magnetic scale, which contributes more and more significantly as $T$ is decreased towards the \emph{phenomenologically relevant regime} where $g\sim{\cal O}(1)$, is beyond the scope of conventional resummed perturbation theory due to the so-called Linde problem~\cite{Linde:1980ts,Gross:1980br}. For heavy-ion experiments to have the greatest possible impact on science, it is essential to make as close as possible a connection to the fundamental theory QCD. There is thus an urgent need for theoretical frameworks based rigorously on QCD which can be applied to the QGP in the phenomenologically relevant regime. In the following, I briefly review the recent developments for this concern from both the electric and magnetic scales based on the HTL effective theory. In Sec.~\ref{sec:electric}, I discuss the progress in the electric scale by a next-to-next-to-leading order calculation of QCD thermodynamics at finite $T$ and $\mu$ from the hard-thermal-loop perturbation theory. In Sec.~\ref{sec:magnetic}, I discuss the progress in the magnetic scale by a first HTL analysis of the collective excitations of hot QCD including the magnetic scale. I conclude in Sec.~\ref{sec:c&o} with an outlook for future perspectives.

\section{Progress in Electric Scale: Thermodynamics}\label{sec:electric}

It had been a longstanding problem that the resulting series of a weak-coupling expansion is poorly convergent at finite $T$ unless the coupling is tiny~\cite{Shuryak:1977ut,Chin:1978gj,Kapusta:1979fh,Toimela:1982hv,Arnold:1994ps,Arnold:1994eb,Zhai:1995ac,Braaten:1995ju,Braaten:1995jr,Kajantie:2002wa}, thus a straightforward perturbative expansion in powers of $g$ for QCD is not of any quantitative use for the temperatures achieved in heavy-ion collisions. The poor convergence stems from the fact that at high $T$ the classical solution is not well-described by massless degrees of freedom, and it is instead better described by massive quasiparticles with non-trivial dispersion relations and interactions generated by the thermal scales. This calls the need for reorganizing the perturbative series which treats the thermal scales more carefully. \emph{Hard-thermal-loop perturbation theory} (HTLpt) is a reorganization scheme for thermal QCD incorporating the electric scale~\cite{Andersen:1999fw}. It systematically shifts the expansion to being around an ideal gas of quasiparticles with a thermal mass on the order of $gT$. HTLpt is a gauge-invariant generalization of \emph{screened perturbation theory}~\cite{Karsch:1997gj,Chiku:1998kd,Andersen:2000yj,Andersen:2001ez,Andersen:2008bz}, which is a reorganization scheme for scalar field theory at finite $T$ inspired in part by \emph{variational perturbation theory}~\cite{Yukalov:1976pm,Stevenson:1981vj,Duncan:1988hw,Duncan:1992ba,Sisakian:1994nn,Janke:1995zz}. The thermodynamic calculations from HTLpt have been firstly carried out at finite $T$ and vanishing $\mu$ at one-loop or leading order (LO)~\cite{Andersen:1999fw,Andersen:1999sf,Andersen:1999va}, two-loop or next-to-leading order (NLO)~\cite{Andersen:2002ey,Andersen:2003zk}, and three-loop or next-to-next-to-leading order (NNLO)~\cite{Andersen:2009tw,Andersen:2009tc,Andersen:2010ct,Andersen:2010wu,Andersen:2011sf,Andersen:2011ug,Su:2011zv}. In recent years, the corresponding calculations have been generalized to finite $T$ and $\mu$~\cite{Andersen:2012wr,Haque:2012my,Haque:2013qta,Mogliacci:2013mca,Haque:2013sja,Haque:2014rua}. The HTLpt framework has been applied to evaluating the equation of state for cold dense quark matter (i.e. vanishing $T$ and finite $\mu$)~\cite{Baier:1999db} with application to compact stars~\cite{Andersen:2002jz}. Application of some HTL motivated approaches to thermodynamics and various susceptibilities can be found in Refs.~\citen{Chakraborty:2001kx,Chakraborty:2002yt,Chakraborty:2003uw,Haque:2011iz,Haque:2010rb,Blaizot:1999ip,Blaizot:1999ap,Blaizot:2000fc,Blaizot:2001vr,Blaizot:2002xz}. In the following of this section, I briefly discuss the setup of HTLpt and the recently obtained NNLO thermodynamic potential at finite $T$ and $\mu$ from which various thermodynamic functions and susceptibilities are derived~\cite{Haque:2014rua}.

\subsection{Hard-Thermal-Loop Perturbation Theory}

Hard-thermal-loop perturbation theory is a reorganization of the perturbative series of thermal gauge theories. The HTLpt Lagrangian density for QCD in Minkowski space can be written as
\be
 {\cal L}=\left.({\cal L}_{\rm QCD}+{\cal L}_{\rm HTL})\right|_{g\rightarrow\sqrt{\delta}g}+\Delta{\cal L}_{\rm HTL} \,.
\label{total_lag} 
\ee
Here ${\cal L}_{\rm QCD}$ is the QCD Lagrangian density that reads
\be
{\cal L}_{\rm QCD}=-\frac{1}{2}{\rm Tr}[G_{\mu\nu}G^{\mu\nu}]+i\bar\psi\gamma^\mu D_{\mu}\psi+{\cal L}_{\rm gh}+{\cal L}_{\rm gf}
+\Delta{\cal L}_{\rm QCD} \, ,
\label{qcd_lag}
\ee
where $G^{\mu\nu}=\partial^{\mu}A^{\nu}-\partial^{\nu}A^{\mu}-ig[A^{\mu},A^{\nu}]$ is the gluon field strength tensor,  $D^{\mu}=\partial^{\mu}-igA^{\mu}$ is the covariant derivative, and the term with the quark fields $\psi$ contains an implicit sum over the $N_f$ quark flavors. The term $\Delta{\cal L}_{\rm QCD}$ contains the counterterms necessary to cancel ultraviolet divergences in perturbative calculations. The ghost term ${\cal L}_{\rm gh}$ depends on the form of the gauge-fixing term ${\cal L}_{\rm gf}$.

The gauge-invariant HTL improvement term ${\cal L}_{\rm HTL}$ reads~\cite{Braaten:1991gm} 
\be
 {\cal L}_{\rm HTL}=(1-\delta)i m_q^2\bar\psi\gamma^\mu\left\langle\frac{y_\mu}{y\cdot\! D}\right\rangle_{\!\hat{\bf y}}\psi-\frac{1}{2}(1-\delta)
 m_D^2 {\rm Tr}\lb G_{\mu\alpha}\left\langle\frac{y^\alpha y_\beta}{(y\cdot\! D)^2}\right\rangle_{\!\hat{\bf y}} G^{\mu\beta}\rb \,,
\nn
\label{htl_lag}
\ee
where $y^\mu = (1, {\bf\hat{y}})$ is a light-like four-vector with ${\bf\hat{y}}$ being a three-dimensional unit vector and the angular bracket indicates an average over the direction of ${\bf\hat{y}}$. The two parameters $m_D$ and $m_q$ can be identified with the Debye screening mass and the quark thermal mass, respectively, and account for screening effects.  HTLpt is defined by treating $\delta$ as a formal expansion parameter. By coupling the HTL improvement term (\ref{htl_lag}) to the QCD Lagrangian (\ref{qcd_lag}), HTLpt systematically shifts the perturbative expansion from being around an ideal gas of massless particles to being around an ideal gas of massive quasiparticles which are more appropriate physical degrees of freedom at high temperature and/or density.

Physical observables are calculated in HTLpt by expanding in powers of $\delta$, truncating at some specified order, and then setting $\delta = 1$. This defines a reorganization of the perturbative series in which the effects of $m_D^2$ and $m_q^2$ terms in (\ref{htl_lag}) are included to leading order but then systematically subtracted out at higher orders in perturbation theory by the $\delta m_D^2$ and $\delta m_q^2$ terms in (\ref{htl_lag}). The HTLpt Lagrangian (\ref{total_lag}) reduces to the QCD Lagrangian (\ref{qcd_lag}) if we set $\delta=1$. If the expansion in $\delta$ could be calculated to all orders, the final result would not depend on $m_D$ and $m_q$ when we set $\delta=1$. However, any truncation of the expansion in $\delta$ produces results that depend on $m_D$ and $m_q$. As a consequence, a prescription is required to determine $m_D$ and $m_q$ as a function of $T$, $\mu$ and $\alpha_s$. Note that HTLpt is gauge invariant order-by-order in the $\delta$ expansion.

The HTLpt expansion generates additional ultraviolet divergences. In QCD perturbation theory, renormalizability constrains the ultraviolet divergences to have a form that can be cancelled by the counterterm Lagrangian $\Delta {\cal L}_{\rm QCD}$. There is yet no proof for the renormalizability of the HTL perturbation expansion, it has been shown in Refs.~\citen{Andersen:2009tw,Andersen:2009tc,Andersen:2010ct,Andersen:2010wu,Andersen:2011sf,Haque:2013sja,Haque:2014rua} that it is possible to renormalize the HTLpt thermodynamic potential through NNLO with a counterterm Lagrangian $\Delta {\cal L}_{\rm HTL}$ containing only a vacuum counterterm, a Debye mass counterterm, a fermion mass counterterm, and a coupling constant counterterm.  The necessary counterterms for renormalization of the NNLO thermodynamic potential are
\be
\Delta {\cal E}_0&=&\frac{d_A}{128\pi^2\epsilon}(1-\delta)^2m_D^4\ ,
\label{ctE}\\
\Delta m_D^2&=&\frac{11c_A-4s_F}{12\pi\epsilon}\alpha_s\delta(1-\delta)m_D^2\ ,
\label{ctmD}
\\
\Delta m_q^2&=&\frac{3}{8\pi\epsilon}\frac{d_A}{c_A}\alpha_s \delta(1-\delta)m_q^2\ ,
\label{ctmq}
\\
\delta\Delta\alpha_s&=&-\frac{11c_A-4s_F}{12\pi\epsilon}\alpha_s^2\delta^2\ ,
\label{ctalpha}
\ee
where $c_A=N_c$, $d_A=N_c^2-1$, $s_F=N_f/2$, $d_F=N_cN_f$, and $s_{2F}=C_F s_f$ with $C_F = (N_c^2-1)/2N_c$. Note that the coupling constant counterterm (\ref{ctalpha}) is consistent with one-loop running of $\alpha_s$.

\subsection{NNLO Thermodynamic Potential at Finite $T$ and $\mu$}

The calculation of the thermodynamic potential in HTLpt involves the evaluation of vacuum diagrams (see Figs. 2 and 3 in Ref.~\citen{Andersen:2011sf} for the diagrams through NNLO).  The fact that $m_D$ and $m_q$ are of the order $gT$ suggests that $m_D/T$ and $m_q/T$ can be treated as expansion parameters of order $g$~\cite{Andersen:2001ez}. In order to make the calculation of the thermodynamic potential analytically tractable in practice, a mass expansion in terms of $m_D/T$ and $m_q/T$ is carried out after the $\delta$ expansion is done. It was shown that the first few terms in the mass expansion gave a surprisingly accurate approximation to the exact result~\cite{Andersen:1999fw,Andersen:1999sf,Andersen:1999va,Andersen:2001ez}. The resulting NNLO thermodynamic potential is completely analytic, and it is accurate to order $g^5$ in the weak-coupling limit. Defining $\aleph(z) \equiv \Psi(z)+\Psi(z^*)$ with $z=1/2-i\hmu$ and $\Psi$ being the digamma function, $\Omega_0 \equiv -d_A\pi^2T^4/45$, and $\hat{x} \equiv x/2\pi T$ for dimensionless variables, the NNLO thermodynamic potential for QCD at finite $T$ and $\mu$ (with $\mu_f$ being the chemical potential for quarks with flavor $f$) reads~\cite{Haque:2014rua}

\begin{eqnarray}
\frac{\Omega_{\rm NNLO}}{\Omega_0}
&=&
\frac{7}{4}\frac{d_F}{d_A}\frac{1}{N_f}\sum\limits_f
\bigg(1+\frac{120}{7}\hmu_f^2+\frac{240}{7}\hmu_f^4\bigg)
- \frac{s_F\alpha_s}{\pi}\frac{1}{N_f}\sum\limits_f
\bigg[\frac{5}{8}\left(1+12\hat\mu_f^2\right)\left(5+12\hat\mu_f^2\right)
\nn
&&
-\,\frac{15}{2}\left(1+12\hat\mu_f^2\right)\hat m_D
- \frac{15}{2}\bigg(2\ln{\frac{\hat\Lambda}{2}-1-\aleph(z_f)}\bigg)\hat m_D^3
+90\hat m_q^2 \hat m_D\bigg]
\nn
&&
+\,\frac{s_{2F}}{N_f}\left(\frac{\alpha_s}{\pi}\right)^2\sum\limits_f
\bigg[\frac{15}{64}
\bigg\{35-32\lb1-12\hmu_f^2\rb\frac{\zeta'(-1)}{\zeta(-1)}+472 \hat\mu_f^2+1328 \hat\mu_f^4
\nn
&&
+\,64\Big(-36i\hat\mu_f\aleph(2,z_f)+6(1+8\hat\mu_f^2)\aleph(1,z_f)
              +3i\hat\mu_f(1+4\hat\mu_f^2)\aleph(0,z_f)\Big)\bigg\}
\nn
&&
-\,\frac{45}{2}\hat m_D\left(1+12\hat\mu_f^2\right)\bigg] 
\nn
&&
+\left(\frac{s_F\alpha_s}{\pi}\right)^2\frac{1}{N_f}\sum\limits_{f}\frac{5}{16}
\Bigg[96\left(1+12\hat\mu_f^2\right)\frac{\hat m_q^2}{\hat m_D}
         +\frac{4}{3}\lb1+12\hmu_f^2\rb\lb5+12\hat\mu_f^2\rb\ln\frac{\hat{\Lambda}}{2}
\nn
&& 
+\,\frac{1}{3}+4\gamma_E+8(7+12\gamma_E)\hat\mu_f^2+112\mu_f^4
    -\frac{64}{15}\frac{\zeta^{\prime}(-3)}{\zeta(-3)}
    -\frac{32}{3}(1+12\hat\mu_f^2)\frac{\zeta^{\prime}(-1)}{\zeta(-1)}
\nn
&&
-\,96\Big\{8\aleph(3,z_f)+12i\hat\mu_f\aleph(2,z_f)-2(1+2\hat\mu_f^2)\aleph(1,z_f)
               -i\hat\mu_f\aleph(0,z_f)\Big\}
\Bigg] 
\nn
&&
+\left(\frac{s_F\alpha_s}{\pi}\right)^2\frac{1}{N_f^2}\sum\limits_{f,g}
\Bigg[\frac{5}{4\hat m_D}\left(1+12\hat\mu_f^2\right)\left(1+12\hat\mu_g^2\right)
         +90\bigg\{ 2\left(1 +\gamma_E\right)\hat\mu_f^2\hat\mu_g^2
\nn
&&
-\,\Big\{\aleph(3,z_f+z_g)+\aleph(3,z_f+z_g^*)
          +4i\hat\mu_f\left[\aleph(2,z_f+z_g)+\aleph(2,z_f+z_g^*)\right]-4\hat\mu_g^2\aleph(1,z_f)
\nn
&&
-\,(\hat\mu_f+\hat\mu_g)^2\aleph(1,z_f+z_g)-(\hat\mu_f-\hat\mu_g)^2\aleph(1,z_f+z_g^*)
   -4i\hat\mu_f\hat\mu_g^2\aleph(0,z_f)\Big\}\bigg\}
\nn
&&
-\,\frac{15}{2}\lb1+12\hat\mu_f^2\rb\bigg(2\L-1-\aleph(z_g)\bigg)  \hat m_D
\Bigg]
\nn
&&
+\left(\frac{c_A\alpha_s}{3\pi}\right)\left(\frac{s_F\alpha_s}{\pi N_f}\right)\sum\limits_f
\Bigg[\frac{15}{2\hat m_D}\lb1+12\hmu_f^2\rb
         -\frac{235}{16}\bigg\{\bigg(1+\frac{792}{47}\hat\mu_f^2
                                                   +\frac{1584}{47}\hat\mu_f^4\bigg)\ln\frac{\hat\Lambda}{2}
\nn
&&
-\,\frac{144}{47}\lb1+12\hmu_f^2\rb\ln\hat m_D
  +\frac{319}{940}\bigg(1+\frac{2040}{319}\hat\mu_f^2+\frac{38640}{319}\hat\mu_f^4\bigg)
  -\frac{24 \gamma_E }{47}\lb1+12\hat\mu_f^2\rb
\nn
&&
    -\,\frac{44}{47}\bigg(1+\frac{156}{11}\hmu_f^2\bigg)\frac{\zeta'(-1)}{\zeta(-1)}
    -\frac{268}{235}\frac{\zeta'(-3)}{\zeta(-3)}
    -\frac{72}{47}\Big[4i\hat\mu_f\aleph(0,z_f)+\left(5-92\hat\mu_f^2\right)\aleph(1,z_f)
\nn
&&
+\, 144i\hmu_f\aleph(2,z_f)
   +52\aleph(3,z_f)\Big]\bigg\}+90\frac{\hat m_q^2}{\hat m_D}
   +\frac{315}{4}\bigg\{\bigg(1+\frac{132}{7}\hmu_f^2\bigg)\L
\nn
&&
+\,\frac{11}{7}\lb1+12\hmu_f^2\rb\gamma_E+\frac{9}{14}\bigg(1+\frac{132}{9}\hmu_f^2\bigg)
+\frac{2}{7}\aleph(z_f)\bigg\}\hat m_D 
\Bigg]
+\frac{\Omega_{\rm NNLO}^{\rm YM}}{\Omega_0} \, ,
\label{finalomega}
\end{eqnarray}
where the sums over $f$ and $g$ include all quark flavors and $\Omega_{\rm NNLO}^{\rm YM}$ is the pure-glue contribution to the thermodynamic potential that reads~\cite{Andersen:2009tc}

\begin{eqnarray}\nonumber
\frac{\Omega_{\rm NNLO}^{\rm YM}}{\Omega_0}
&=&
1-{15\over4}\hat{m}_D^3
+{N_c\alpha_s\over3\pi}
\bigg[
-{15\over4}+{45\over2}\hat{m}_D-{135\over2}\hat{m}^2_D
-{495\over4}\bigg(\L+{5\over22}+\gamma_E\bigg)\hat{m}^3_D
\bigg]
\nn 
&&
+\left({N_c\alpha_s\over3\pi}\right)^2
\bigg[{45\over4\hat{m}_D}-{165\over8}
\bigg(\L-{72\over11}\ln{\hat{m}_D}-{84\over55}-{6\over11}\gamma_E
         -{74\over11}{\zeta^{\prime}(-1)\over\zeta(-1)}
\nn
&&
+\,{19\over11}{\zeta^{\prime}(-3)\over\zeta(-3)}\bigg)
+{1485\over4}\bigg(\L-{79\over44}+\gamma_E+\log2-{\pi^2\over11}\bigg)\hat{m}_D
\bigg]
\,.
\label{Omega-NNLO}
\end{eqnarray}

As discussed in Ref.~\citen{Andersen:2011sf}, the two-loop perturbative electric mass for gluons introduced by Braaten and Nieto in Refs.~\citen{Braaten:1995ju,Braaten:1995jr} originally at finite $T$ and vanishing $\mu$ is the most suitable one for NNLO HTLpt calculations, and it will be thus adopted in the next sections. The finite $T$ and $\mu$ generalization was obtained in Ref.~\citen{Vuorinen:2003fs} and the resulting $m_D^2$ reads
\begin{eqnarray}
\hat m_D^2&=&\frac{\alpha_s}{3\pi} \Biggl\{c_A
+\frac{c_A^2\alpha_s}{12\pi}\lb5+22\gamma_E+22\Lg\rb +
\frac{1}{N_f} \sum\limits_{f}
\Biggl[ s_F\lb1+12\hmu_f^2\rb
\nonumber\\
  &&+\frac{c_As_F\alpha_s}{12\pi}\lb\lb9+132\hmu_f^2\rb+22\lb1+12\hmu_f^2\rb\gamma_E+2\lb7+132\hmu_f^2\rb\L+4\aleph(z_f)\rb
\nonumber\\  
&&+\frac{s_F^2\alpha_s}{3\pi}\lb1+12\hmu_f^2\rb\lb1-2\L+\aleph(z_f)\rb
 -\frac{3}{2}\frac{s_{2F}\alpha_s}{\pi}\lb1+12\hmu_f^2\rb \Biggr] \Biggr\} \, .
\end{eqnarray}
The effect of the in-medium quark mass parameter $m_q$ in thermodynamic functions is small and it is thus set to 0 which is the three-loop variational solution for $m_q$ following Ref.~\citen{Andersen:2011sf}.

\subsection{Thermodynamic Functions}

In this section I review results for the NNLO HTLpt pressure, trace anomaly, and speed of sound obtained in Refs.~\citen{Andersen:2011sf,Haque:2014rua}. In all the results, the one-loop running corresponding to~Eq.~\ref{ctalpha} of the coupling renormalization is used. It reads
\be
\alpha_s(\Lambda)&=&\frac{1}{b_0 t} \,.
\ee
with $t = \ln(\Lambda^2/\Lambda_{\overline{\rm MS}}^2)$ and $b_0 = (11c_A-2N_f)/12\pi$. The scale $\Lambda_{\overline{\rm MS}}$ is fixed by requiring that $\alpha_s({\rm 1.5\,GeV}) = 0.326$ which is obtained from lattice measurements~\cite{Bazavov:2012ka}. This gives $\Lambda_{\overline{\rm MS}} = 176$\,MeV. For the renormalization scale, $\Lambda_g$ and $\Lambda_q$ are used for purely-gluonic and fermionic graphs, respectively.  The central values of these renormalization scales are taken to be $\Lambda_g = 2\pi T$ and $\Lambda=\Lambda_q=2\pi \sqrt{T^2+\mu^2/\pi^2}$.  In all plots the thick lines indicate the result obtained using these central values and the light-blue bands indicate the variation of the results under variation of both of these scales by a factor of 2, e.g. $\pi T \leq \Lambda _g \leq 4 \pi T$.  For all numerical results below $c_A = N_c=3$ and $N_f=3$ are set.

\subsubsection{Pressure}

\begin{figure}[t]
\subfigure{
\hspace{-2mm}\includegraphics[width=0.49\textwidth]{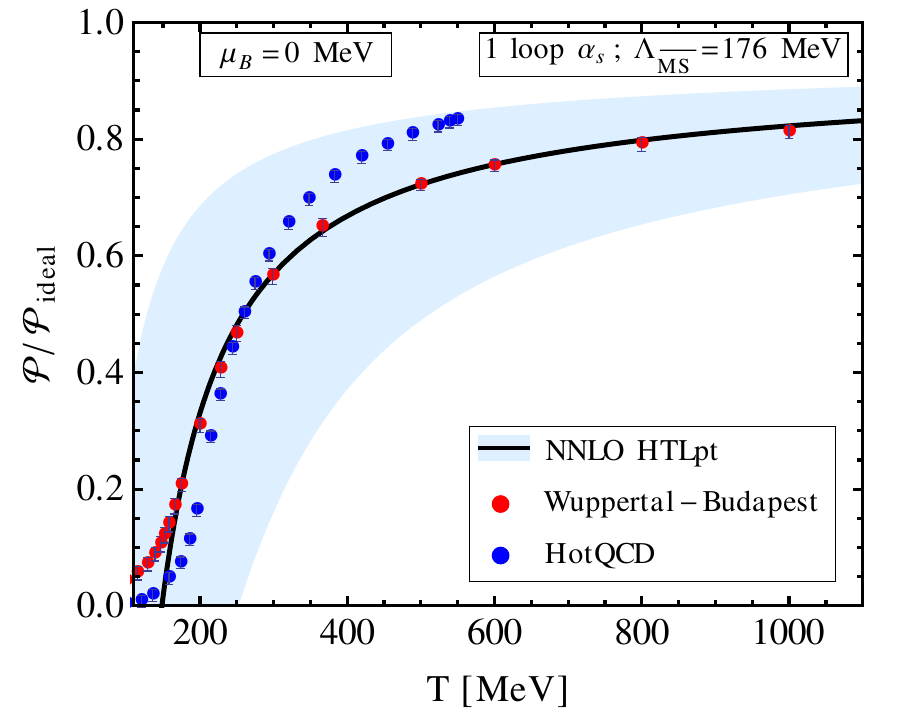}} 
\subfigure{
\includegraphics[width=0.49\textwidth]{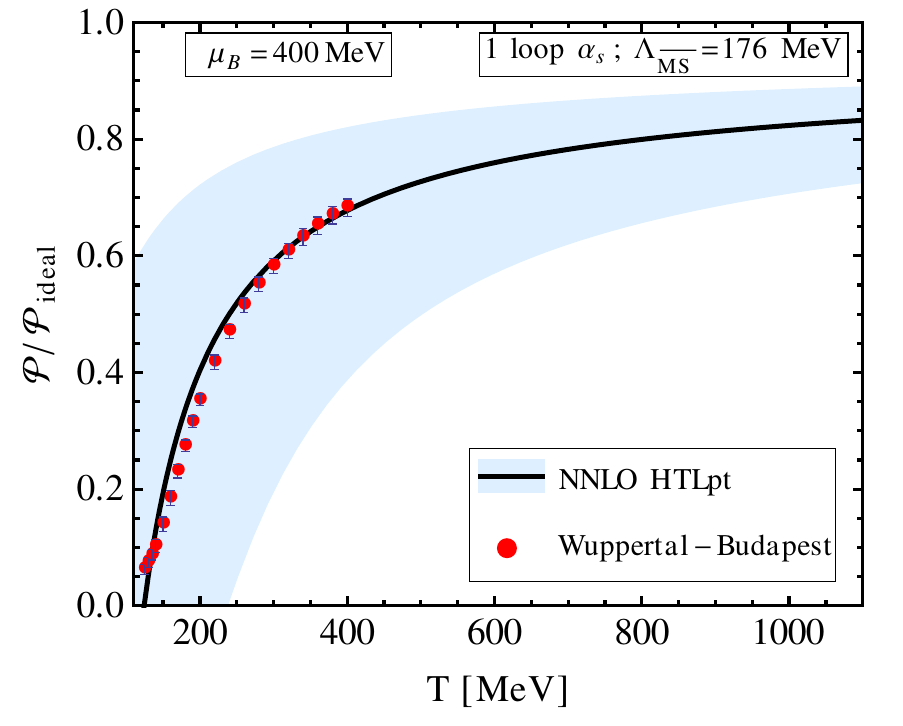}}
\caption{
Comparison of the $N_f=2+1$, $\mu_B=0$ (left) and $\mu_B=400$ MeV (right) NNLO HTLpt pressure with lattice data from Borsanyi et al.~\cite{Borsanyi:2010cj,Borsanyi:2012cr} and Bazavov et al.~\cite{Bazavov:2009zn}.
}
\label{pres_1l}
\end{figure}

The pressure is a key quantity from which all other thermodynamic functions can be derived. It is obtained directly from the thermodynamic potential by
\be
{\cal P} \,=\, -\Omega_{\rm NNLO} \,.
\ee
Fig.~\ref{pres_1l} shows the scaled NNLO HTLpt pressure for $\mu_B=0$ (left) and $\mu_B=400$\,MeV (right) with lattice data from Refs.~\citen{Bazavov:2009zn,Borsanyi:2010cj,Borsanyi:2012cr}. As we can see from this figure, there is quite good agreement between the NNLO HTLpt pressures and the lattice data for $T\gtrsim200$\,MeV when the central value of the scale is used. Since HTLpt does not incorporate the center symmetry $Z(N_c)$, there is no reason to expect agreement with the lattice data at temperatures close to $T_c$ and the agreement seen may be fortuitous.

\subsubsection{Trace Anomaly}

\begin{figure}[t]
\subfigure{
\hspace{-2mm}
\includegraphics[width=0.49\textwidth]{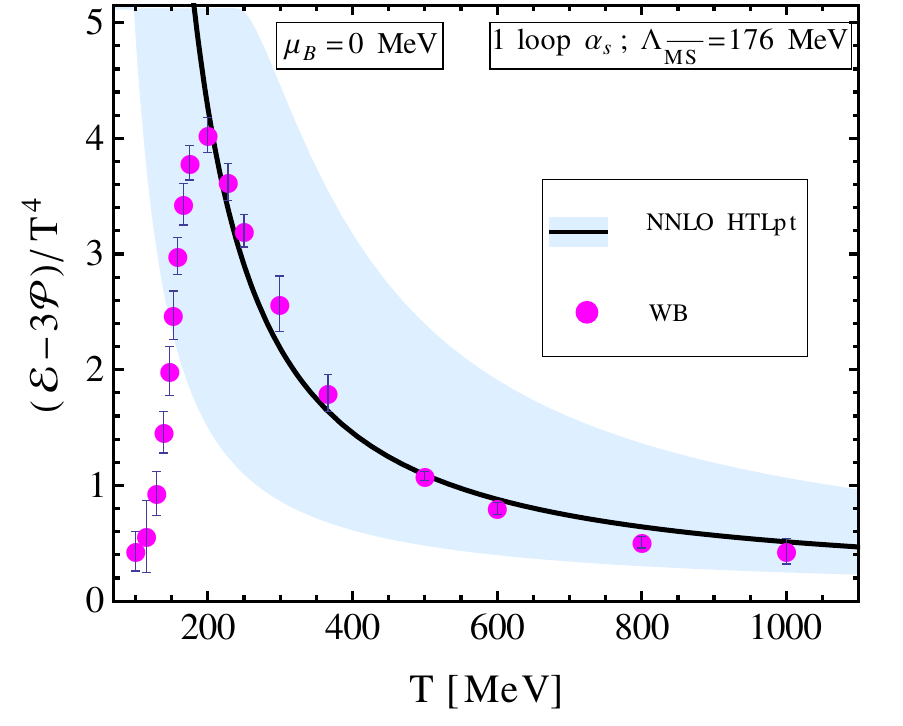}} 
\subfigure{
\includegraphics[width=0.49\textwidth]{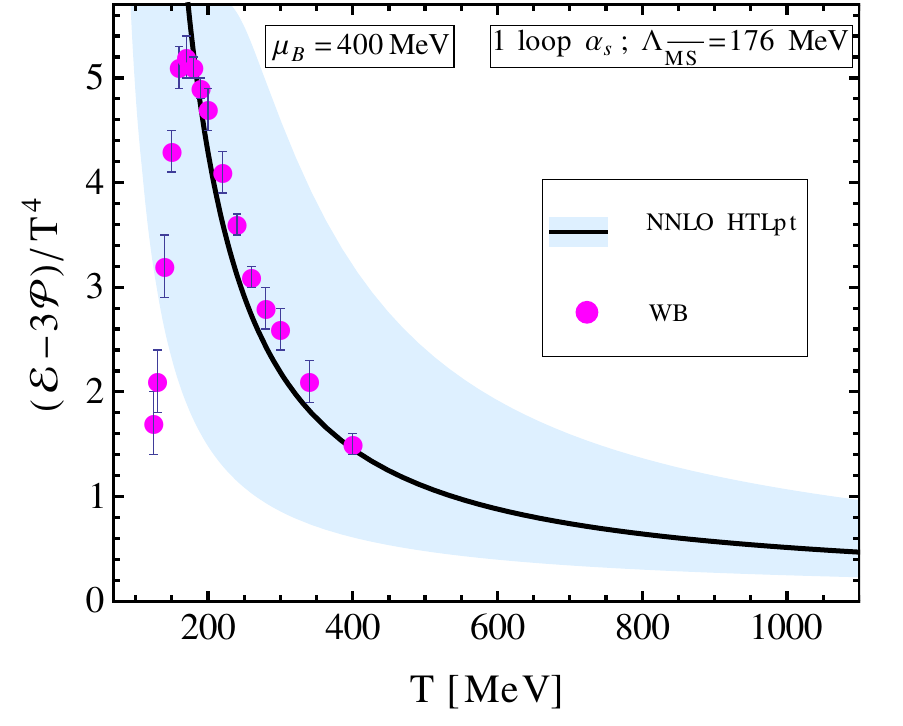}} 
\caption{
Comparison of the $N_f=2+1$, $\mu_B=0$ (left) and $\mu_B=400$ MeV (right) NNLO HTLpt trace anomaly with lattice data. The $\mu_B=0$ lattice data are from Ref.~\citen{Borsanyi:2010cj} and the $\mu_B=400$\,MeV lattice data are from Ref.~\citen{Borsanyi:2012cr}.}
\label{ta_1l}
\end{figure}

The trace anomaly is obtained from the pressure by
\be
{\cal E} - 3 {\cal P} \,=\, T^5 \frac{d}{dT}\left( \frac{{\cal P}}{T^4} \right) \,.
\ee
This quantity is the trace of the energy-momentum tensor and vanishes for an ideal gas of massless particles, it thus measures the breaking of conformal symmetry by quantum effects for such a system. Fig.~\ref{ta_1l} shows the NNLO HTLpt trace anomaly scaled by $T^4$ for $\mu_B = 0$ (left) and $\mu_B = 400$\,MeV (right) together with lattice data from Refs.~\citen{Borsanyi:2010cj,Borsanyi:2012cr}. As we can see from this figure, there is quite good agreement between the NNLO HTLpt trace anomalies and the lattice data for $T\gtrsim220$\,MeV when the central value of the scale is used.

\subsubsection{Speed of Sound}

\begin{figure}[t]
\subfigure{
\hspace{-4mm}
\includegraphics[width=0.49\textwidth]{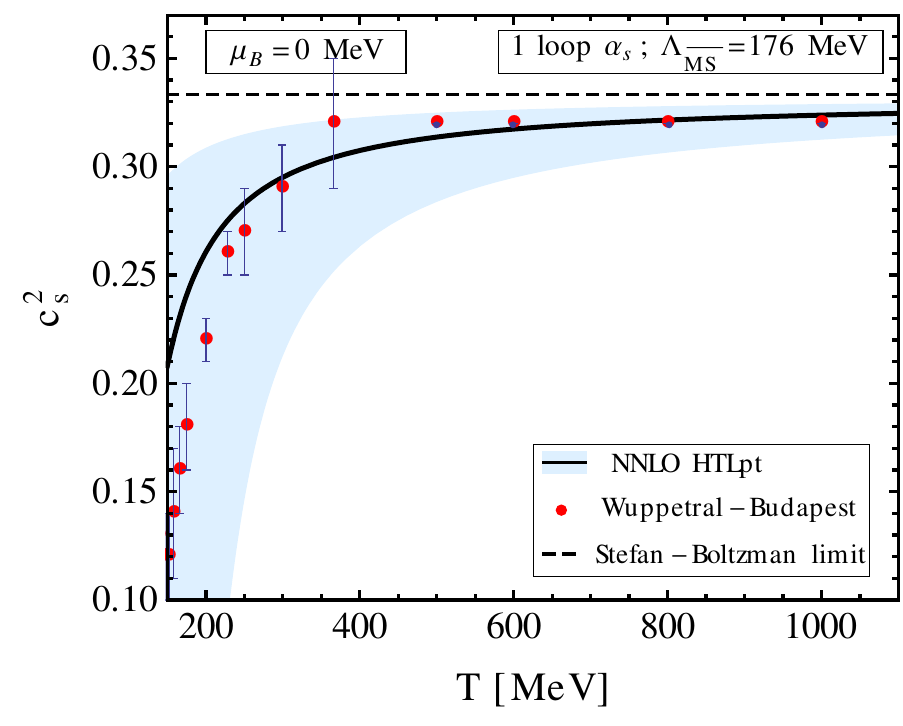}} 
\subfigure{
\includegraphics[width=0.49\textwidth]{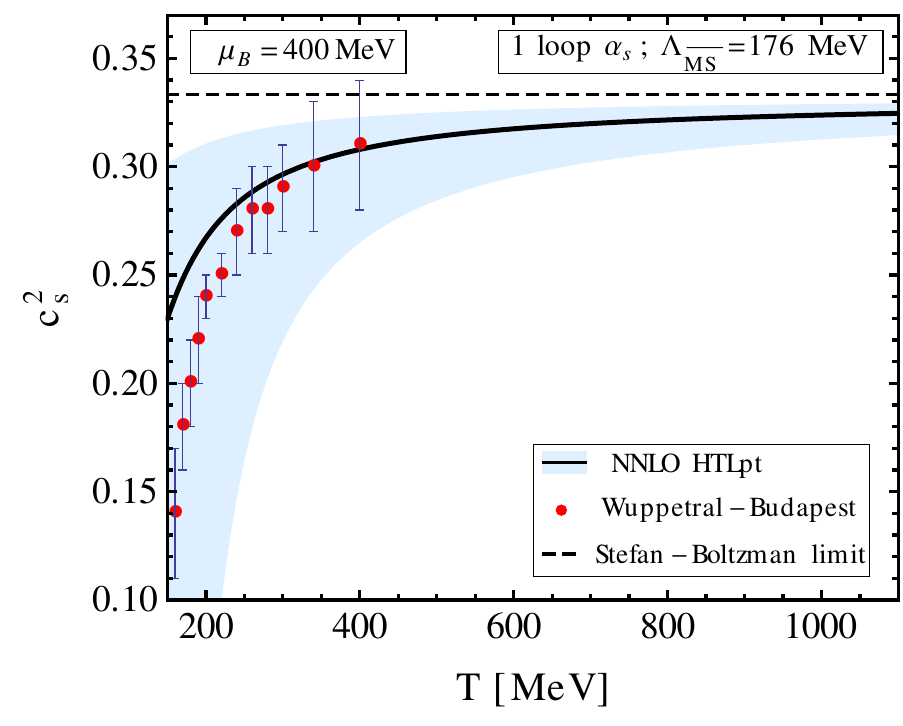}} 
\caption{
Comparison of the $N_f=2+1$, $\mu_B=0$ (left) and $\mu_B=400$\,MeV (right) NNLO HTLpt 
speed of sound with lattice data. The $\mu_B=0$ lattice data are from Ref.~\citen{Borsanyi:2010cj} and the $\mu_B=400$\,MeV lattice data are from Ref.~\citen{Borsanyi:2012cr}.
}
\label{cssq_1l}
\end{figure}

The speed of sound $c_s$ is another phenomenologically relevant quantity defined as
\be
c_s^2 \,=\, \frac{\del{\cal P}}{\del{\cal E}} \,.
\ee
Fig.~\ref{cssq_1l} shows the NNLO HTLpt speed of sound for $\mu_B = 0$ (left) and $\mu_B = 400$\,MeV (right) together with lattice data from Refs.~\citen{Borsanyi:2010cj,Borsanyi:2012cr}. As we can see from this figure, there is quite good agreement between the NNLO HTLpt speeds of sound and the lattice data when the central value of the scale is used.

\subsection{Susceptibilities}

\begin{figure}[t]
\subfigure{
\hspace{-2mm}
\includegraphics[width=0.49\textwidth]{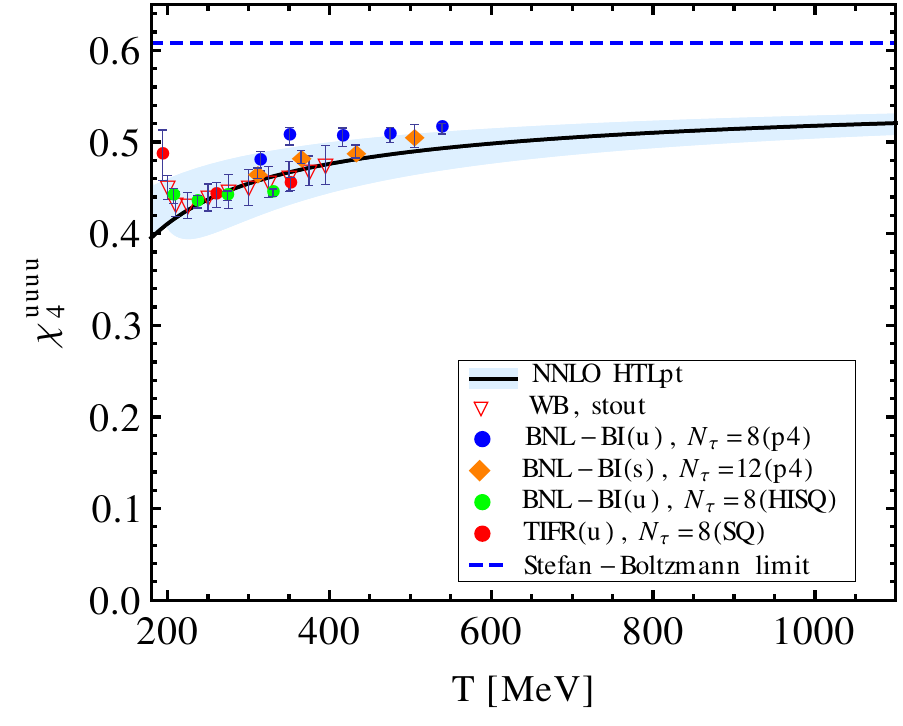}}
\subfigure{
\includegraphics[width=0.49\textwidth]{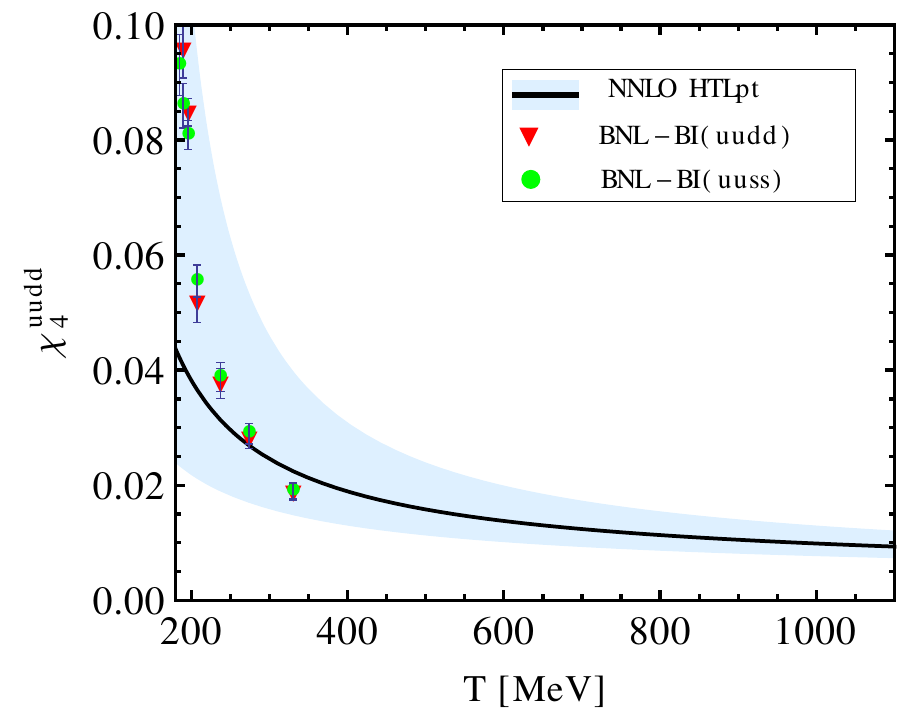}}
\caption{
Comparison of the $N_f=2+1$ NNLO HTLpt results of the 4th order diagonal single quark number susceptibility (left) and the only non-vanishing 4th order off-diagonal quark number susceptibility (right) with lattice data. In the left figure the dashed blue line indicates the Stefan-Boltzmann limit for this quantity. The data labeled BNL-BI(uudd), BNL-BI(u,s), BNL-BI(uuss), and TIFR come from Refs.~\citen{Bazavov:2013dta}, \citen{Bazavov:2013uja}, \citen{Bazavov:2012vg}, and \citen{Datta:2014zqa}, respectively.
}
\label{figsingleq}
\end{figure}

The quark number susceptibilities are another set of phenomenologically relevant quantities. These functions carry information about the response of the system to nonzero density. By taking derivatives of the pressure with respect to chemical potentials, we obtain the quark number susceptibilities
\be
\chi_{ijk\,\cdots}\left(T\right)
\,&\equiv&\,
\left.\frac{\partial^{i+j+k+ \, \cdots}\; {\cal P}\left(T,\bm{\mu}\right)}{\partial\mu_u^i\, \partial\mu_d^j \, \partial\mu_s^k\, \cdots} \right|_{\bm{\mu}=0} \,,
\label{qnsdef}
\ee
where $\bm{\mu}\equiv(\mu_u,\mu_d,...,\mu_{N_f})$ representing a separate chemical potential for each quark flavor. The left panel of Fig.~\ref{figsingleq} shows the 4th order single quark susceptibility $\chi_4^{\rm uuuu}$ comparing to lattice data from Refs.~\citen{Bazavov:2013dta,Bazavov:2013uja,Bazavov:2012vg,Datta:2014zqa}. As can be seen from this figure, the scale variation of the HTLpt result is quite small for this particular quantity and there is good agreement with the lattice data. The right panel of Fig.~\ref{figsingleq} shows the 4th order off-diagonal single quark susceptibility $\chi_4^{\rm uudd}$, which is also in reasonably good agreement with the lattice data.

\begin{figure}[t]
\subfigure{
\hspace{-2mm}
\includegraphics[width=0.49\textwidth]{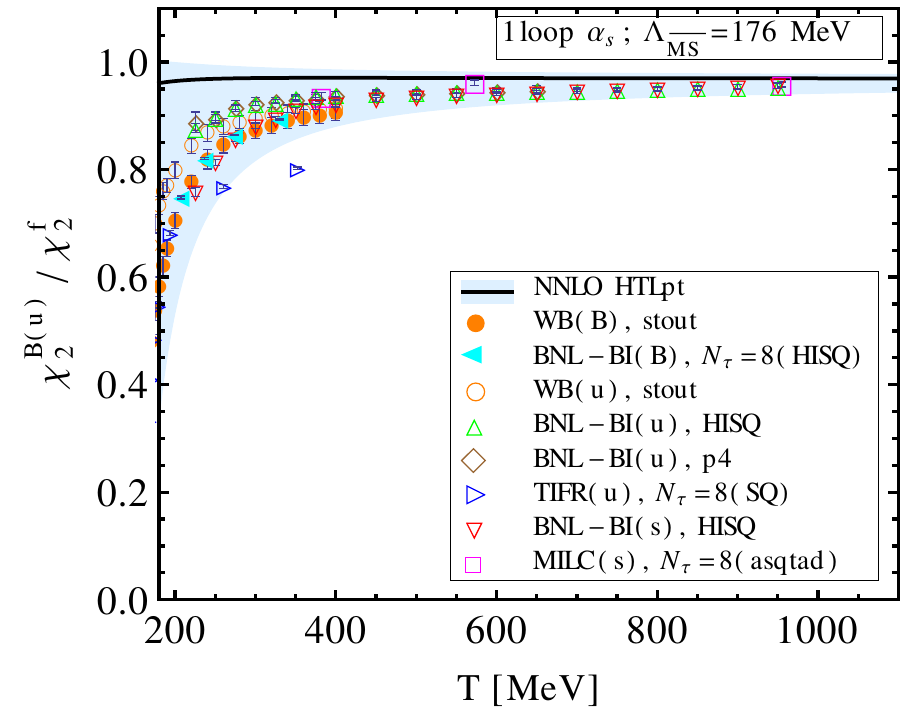}}
\subfigure{
\includegraphics[width=0.49\textwidth]{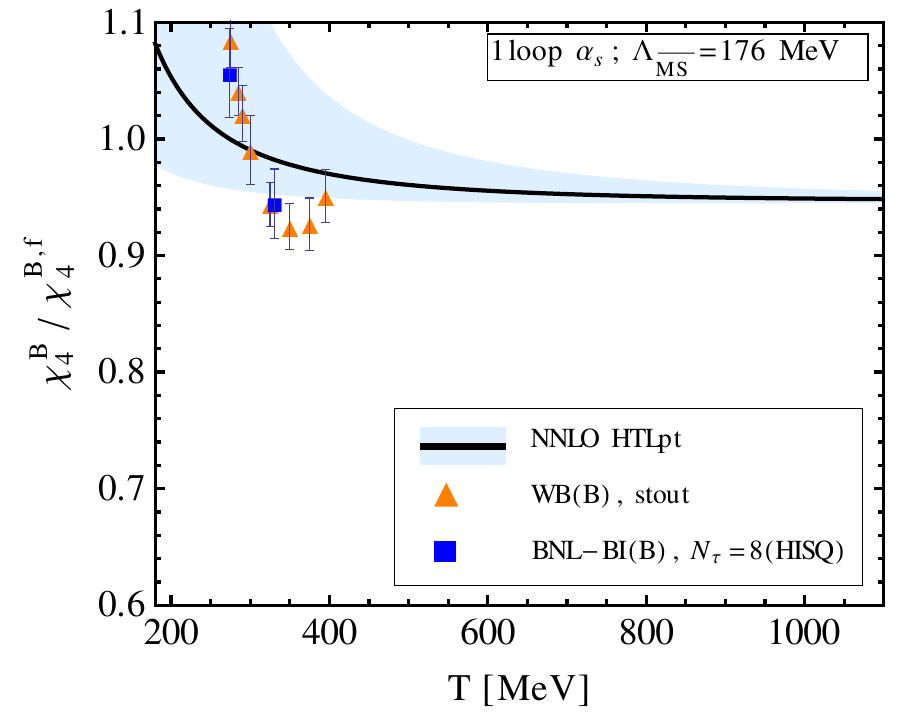}}
\caption{
The scaled 2nd (left) and 4th (right) order baryon number susceptibilities compared with various lattice data. The lattice data labeled WB, BNL-BI(B), BNL-BI(u,s), MILC, and TIFR come from Refs.~\citen{Borsanyi:2011sw}, \citen{Bazavov:2013dta}, \citen{Bazavov:2013uja}, \citen{Bernard:2004je}, and \citen{Datta:2014zqa}, respectively.
}
\label{qns_1l}
\end{figure}

Since the directly accessible information in the experiments are baryon number fluctuations rather than quark number fluctuations, we should pay particular attention to the baryon number susceptibilities defined as
\be
\chi_B^n(T) \equiv \left.\frac{\partial^n {\cal P}}{\partial \mu_B^n}\right|_{\mu_B=0} \,,
\ee
with $\mu_B=\mu_u+\mu_d+\mu_s$. The left panel of Fig.~\ref{qns_1l} shows the scaled 2nd order baryon number susceptibility comparing to lattice data from Refs.~\citen{Borsanyi:2011sw,Bazavov:2013dta,Bazavov:2013uja,Bernard:2004je,Datta:2014zqa}. As can been seen from this figure, the NNLO HTLpt result is in good agreement with the lattice data for $T\gtrsim300$\,MeV. The right panel of Fig.~\ref{qns_1l} shows the scaled 4th order baryon number susceptibility comparing to the lattice data. The NNLO HTLpt result is consistent with the lattice data shown, however the lattice error bars on this quantity are somewhat large and the data are restricted to $T\lesssim400$\,MeV, making it difficult to draw firm conclusions from this comparison. That being said, HTLpt makes a clear prediction for the $T$-dependence of the 4th order baryon number susceptibility. It will be very interesting to see if future lattice data agree with this prediction.

\section{Progress in Magnetic Scale: Collective Excitations}\label{sec:magnetic}

As mentioned in Sec.~\ref{sec:intro}, conventional thermal perturbation theory breaks down at the magnetic scale $g^2T$ due to the Linde problem~\cite{Linde:1980ts,Gross:1980br}. The nonperturbative nature of the magnetic scale is intimately related to the confining property of the dimensionally reduced Yang-Mills theory at high temperature. This suggests that a confinement mechanism should be incorporated within perturbative expansions even when dealing with the deconfined QGP phase. There have been many efforts over the last two decades trying to reconcile resummed perturbation theory with the magnetic scale and this is still a key open question in the field~\cite{Buchmuller:1994qy,Alexanian:1995rp,Jackiw:1995nf,Jackiw:1997jga,Cornwall:1997dc,Eberlein:1998yk,Bieletzki:2012rd}. Color confinement is deeply related to \emph{positivity violation} of the spectral function: if the spectral function of a particle is not positive semi-definite, no K{\"a}ll{\'e}n-Lehmann representation exists, it is then not part of the physical spectrum and thus confined (see Ref.~~\citen{Alkofer:2000wg} for a review). This problem has been studied intensively using lattice QCD and functional methods at both zero and finite temperatures for gluons (see Refs.~\citen{Cucchieri:2004mf,Bowman:2007du,Silva:2014sea,Alkofer:2003jj,Maas:2004se,Maas:2005hs,Fischer:2008uz,Fister:2011uw,Strauss:2012dg} and references therein, see also Refs.~\citen{Maas:2005ym,Maas:2011se} for reviews). The study of the quark sector has not been equally conclusive.

Conventional thermal field approaches to hot QCD are based on massive quasiparticles which only generate short-range correlations. In order to describe a strongly coupled QGP, long-range correlations, whose carriers are light and/or massless modes, are a crucial ingredient. There have been hints on massless modes in a QGP from functional methods~\cite{Harada:2008vk,Qin:2010pc,Nakkagawa:2011ci,Nakkagawa:2012ip,Gao:2014rqa}. At the thermal field frontier, there has been a series of studies on massless modes in Nambu--Jona-Lasinio model, Yukawa model, QED and QCD~\cite{Kitazawa:2005mp,Kitazawa:2006zi,Kitazawa:2007ep,Hidaka:2011rz,Satow:2013oya,Blaizot:2014hka}. In the following, I briefly review the first study on massless modes in hot QCD using confining gluons recently reported in Ref.~\citen{Su:2014rma} which shows genuine non-Abelian features such as positivity violation.

\subsection{Gribov-Zwanziger Formalism at Finite $T$}

A formalism to tackle the issue is the Gribov-Zwanziger (GZ) action, which is well-known from the study of color confinement~\cite{Gribov:1977wm,Zwanziger:1989mf}. It regulates the IR behavior of QCD by fixing the residual gauge transformations, i.e., Gribov copies (see Ref.~\citen{Dokshitzer:2004ie} for a review), that remain after applying the Faddeev-Popov procedure. The GZ action is renormalizable, and it thus provides a systematic framework for perturbative calculations (i.e., $g\ll1$) incorporating confinement effects. The gluon propagator in general covariant gauge reads
\be
D^{\mu\nu}(P)
\,=\,
\left[ \delta^{\mu\nu} - (1-\xi)\frac{P^\mu P^\nu}{P^2} \right] \frac{P^2}{P^4 + \gamma_\smallG^4}
\;,
\label{eq:gluon}
\ee
where $\xi$ is the gauge parameter and the Landau gauge $\xi=0$ has been well explored in practice (see Refs.~\citen{Sobreiro:2005ec,Vandersickel:2012tz} for reviews). The Gribov parameter $\gamma_\smallG$ is solved self-consistently from a gap equation that is defined to infinite loop orders. The GZ gluon propagator is IR suppressed, manifesting confinement effects, and it is a significant improvement over the one from the Faddeev-Popov quantization which forms the basis for conventional perturbative calculations. The gap equation at one-loop order can be solved analytically at asymptotically high $T$ and gives~\cite{Zwanziger:2006sc,Fukushima:2013xsa}
\be
\gamma_\smallG
\,=\,
\frac{D-1}{D} \frac{N_c}{4\sqrt{2}\pi} g^2T
\;,
\label{eq:gamma}
\ee
where $D$ is the space-time dimensions. Eq.~(\ref{eq:gamma}) provides a fundamental IR cutoff at the magnetic scale for the finite-$T$ GZ action. The effectiveness of the GZ framework in the study of Yang-Mills thermodynamics has been explored in Refs.~\citen{Zwanziger:2004np,Zwanziger:2006sc,Lichtenegger:2008mh,Fukushima:2013xsa}.

\subsection{Quark Thermal Self-Energy}

An important measure for the collective behavior of a QGP is the self-energy of quarks and gluons, from which thermal masses, dispersion relations, and spectral functions of collective excitations are derived. The Euclidean one-loop quark self-energy reads 
\be
\Sigma(P)
\,=\,
(ig)^2 C_F \sumint_{\!\!\!\!\{ K \}} \gamma^\mu S(K) \gamma^\nu D^{\mu\nu}(P-K)
\;,
\label{eq:Sigma1}
\ee
where $S(P)=1/{/\!\!\!\!P}$ is the quark propagator, and $D^{\mu\nu}(P)$ is the gluon propagator which is taken from Eq.~(\ref{eq:gluon}). It is worth noting that there have been similar studies for the quark self-energy with nonperturbative gluons at finite density~\cite{Kojo:2009ha,Kojo:2011cn} and in strong magnetic fields~\cite{Kojo:2012js}.

When $g \ll 1$, the leading contribution from $\Sigma(P)$ to the two-point correlation function is from soft external momenta $P \sim gT$, and the leading contribution to the loop integral in Eq.~(\ref{eq:Sigma1}) is from $k \sim T$~\cite{Braaten:1989mz}. This suggests that for studying the high-$T$ behavior of the self-energy in the small-coupling regime, we may expand Eq.~(\ref{eq:Sigma1}) in terms of small $P$ following the systematics of the HTL effective theory. As a result, the gauge-invariant contribution to Eq.~(\ref{eq:Sigma1}) reads~\cite{Su:2014rma}
\be
\Sigma(P) 
&\simeq& 
-(ig)^2 C_F 
\sum_\pm \int_0^\infty\frac{\d k}{2\pi^2}k^2 \int\frac{\d\Omega}{4\pi} \frac{\tilde n_\pm(k,\gamma_\smallG)}{4E_\pm^0} 
\nn &&
\times
\left[
\frac{i\gamma_0 + \kh \cdot{\boldsymbol\gamma}}{iP_0 + k - E_\pm^0 + \frac{ \p \cdot \k}{E_\pm^0}}
+
\frac{i\gamma_0 - \kh \cdot{\boldsymbol\gamma}}{iP_0 - k + E_\pm^0 - \frac{ \p \cdot \k}{E_\pm^0}}
\right]
\;,
\label{eq:Sigma2}
\ee
where $\kh = \k/k$ with $k= |\k|$,  $E_\pm^0=\sqrt{k^2 \pm i\gamma_\smallG^2}$, $\tilde n_\pm(k,\gamma_\smallG)\equiv n_B(\sqrt{k^2 \pm i \gamma_\smallG^2}) + n_F(k)$ with $n_B$ and $n_F$ the Bose-Einstein and Fermi-Dirac distributions, and $\int \d\Omega = \int_0^{2\pi}\d\phi \int_0^\pi \d\cos\theta$.

\subsection{Quark Thermal Mass}

\begin{figure}[t]
\centerline{\includegraphics[width=0.6\textwidth]{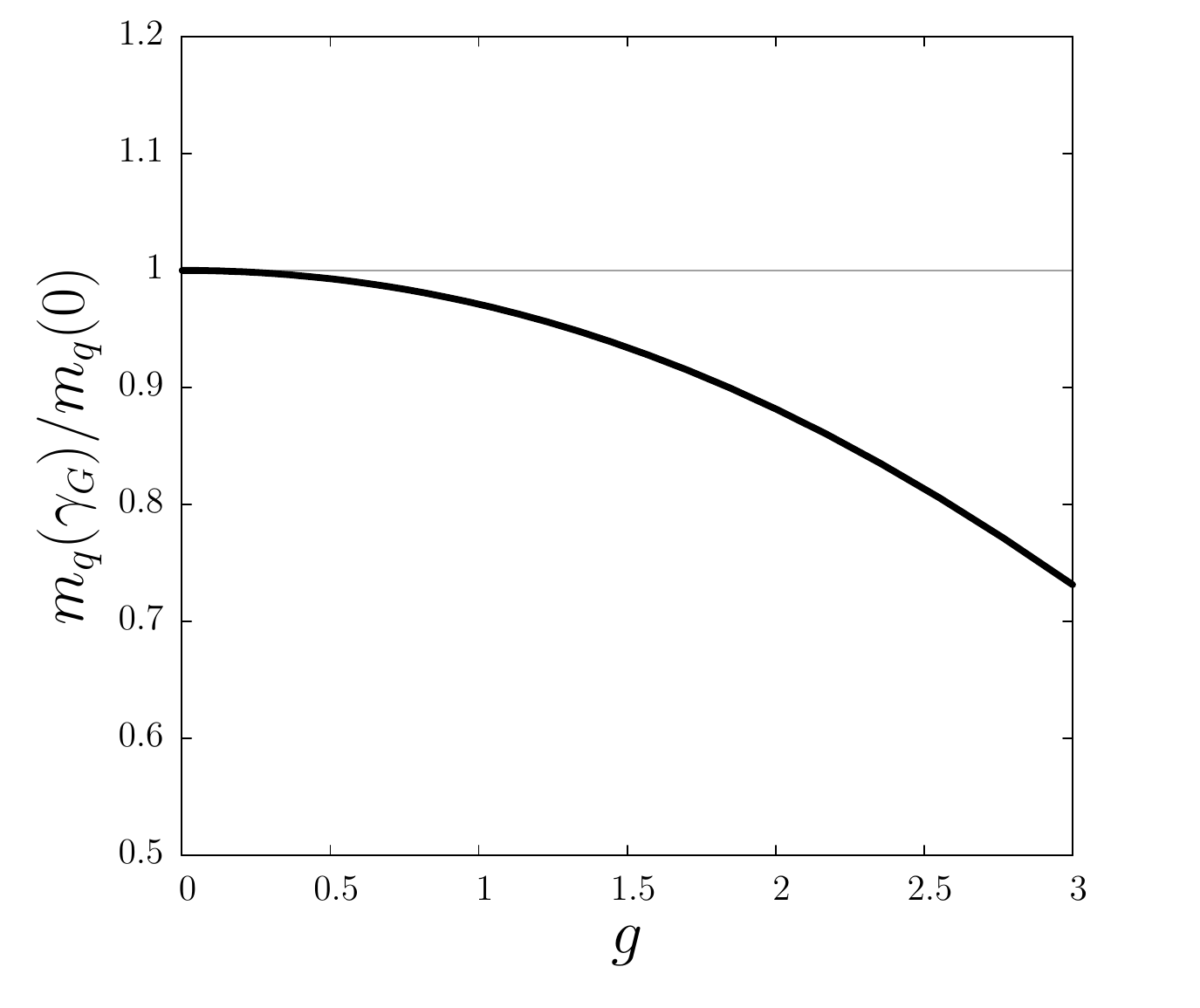}}
\caption{The quark thermal mass $m_q(\gamma_\smallG)$ from Eq.~(\ref{eq:mq1}) scaled by the perturbative value $m_q(0)$.}
\label{fig:mq}
\end{figure}

The quark thermal mass incorporating effects from the magnetic scale reads
\be
m_q^2(\gamma_\smallG)
\,=\,
\frac{g^2 C_F}{4\pi^2} 
\sum_\pm \int_0^\infty \!\! dk \frac{k^2 \tilde n_\pm(k,\gamma_\smallG)}{E_\pm^0} \,,
\label{eq:mq1}
\ee
which reduces to the conventional HTL one, $m_q^2(0) =C_F g^2T^2/8$, for $\gamma_\smallG= 0$. The scaled quark thermal mass $m_q(\gamma_\smallG)/m_q(0)$ is shown in Fig.~\ref{fig:mq}. It is clear from the figure that $m_q$ receives negative contributions from $\gamma_\smallG$, which is a manifestation of anti-screening effects generated by the magnetic scale. Although the effect is modest in the studied range of couplings, this is a profound signal of the build-up of long-range correlations in the system and similar anti-screening effects have been observed on the lattice for the Debye screening mass~\cite{Kaczmarek:2005ui}.

\subsection{Massless Mode and Positivity Violation}

\begin{figure}[t]
\centerline{\includegraphics[width=0.75\textwidth]{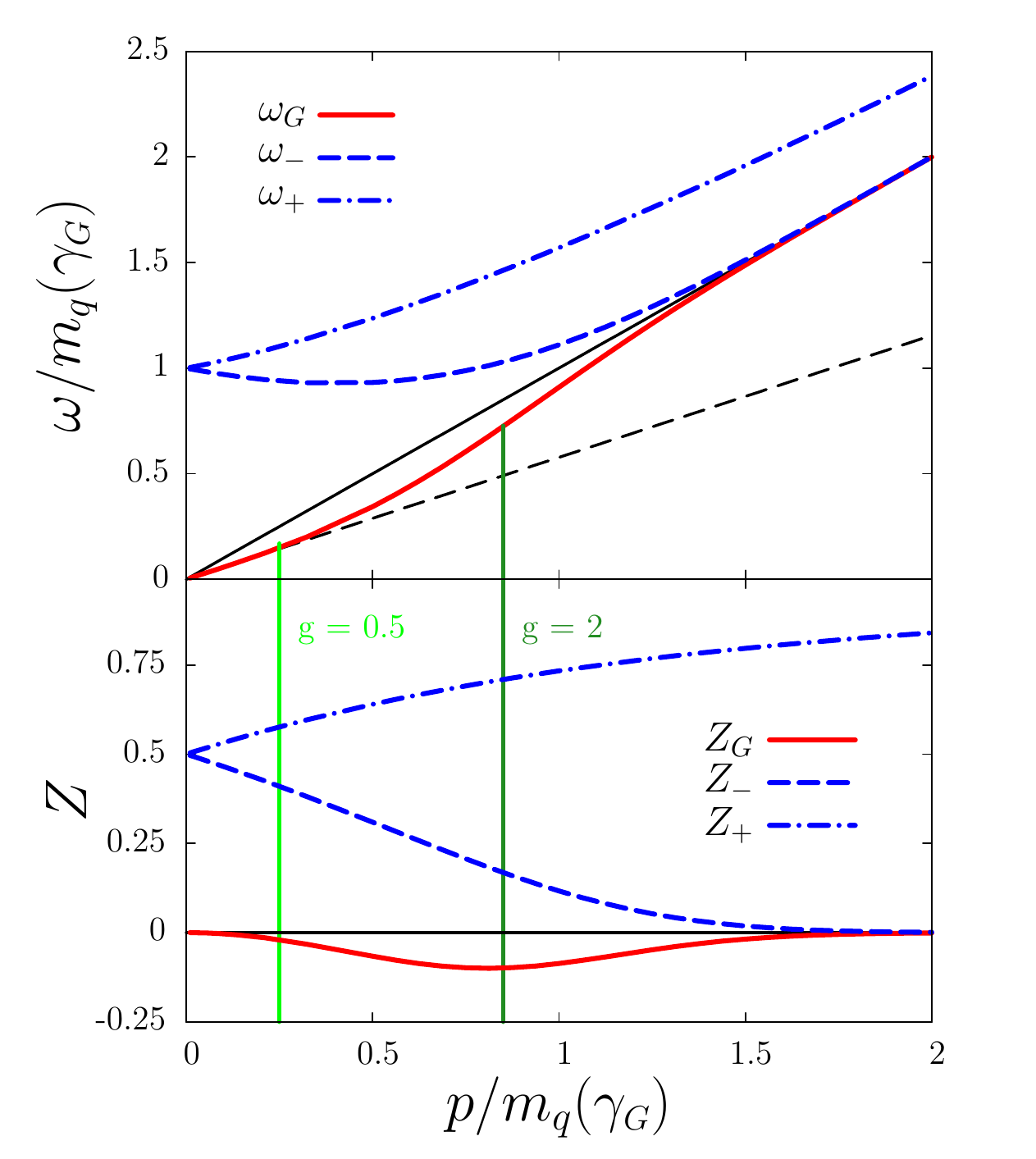}}
\caption{Dispersion relations (upper panel) and the corresponding residues (lower panel) for the particle ($\omega_+$), plasmino ($\omega_-$) and Gribov ($\omega_G$) poles.}
\label{fig:disp}
\end{figure}

The dispersion relation is obtained by analytically continuing the self-energy Eq.~(\ref{eq:Sigma2}) to Minkowski space and then solving the poles in the corresponding quark propagator $iS^{-1}(P) = \slashed P - \Sigma(P) = 0$. The resulting dispersion relations and residues of the poles are displayed in the upper and lower panels of Fig.~\ref{fig:disp}. 

In contrast to the conventional HTL case, there are three poles in the propagator. Firstly, the screened quasi-particle excitations are recovered,
\be
\omega = \omega_+(p; \gamma_\smallG)\,, \hspace{2em} \omega = \omega_-(p;\gamma_\smallG) \,,
\ee
the so-called particle $\omega_+$ and plasmino $\omega_-$ modes, with $\omega_\pm (0; \gamma_\smallG) = m_q(\gamma_\smallG)$ as expected. Both $\omega_\pm/m_q(\gamma_\smallG)$ and $Z_\pm$ are $g$-independent, and this has been verified explicitly up to $g\sim2$ in Ref.~\citen{Su:2014rma}. This property is exactly the same as in the conventional HTL effective theory, and it is thus a non-trivial consistency check of the setup.

In addition to the massive modes, there exists a novel excitation named \emph{Gribov pole} as in Ref.~\citen{Su:2014rma},
\be
\omega = \omega_\smallG(p; \gamma_\smallG) \,.
\ee
It describes \emph{massless} fermionic excitations in the plasma with dispersion relation $\omega = v_s p$ at small momenta, with $v_s \approx 1/\sqrt{3}$ (speed of sound) independent of $g$ for the studied range. The Gribov mode ``grows'' in the $(\omega, p)$-plane while the magnetic scale is increasing (through increasing $g$), and this effectively introduces a new \emph{magnetic scaling} behavior to the non-Abelian plasma. The vertical lines in Fig.~\ref{fig:disp} schematically demonstrate how the Gribov mode grows: at small coupling, e.g., $g=0.5$, the mode terminates at rather small momentum; as the coupling increases, to e.g., $g=2$, the permitted momentum range increases accordingly. At larger momenta than the permitted ones for each coupling, we are hitting branch cuts and Landau damping takes place as a consequence. The Gribov pole goes along with a residue $Z_G(p) < 0$ which directly implies \emph{positivity violation} of the corresponding spectral functions in the region of space-like momenta. These novel features are direct manifestations of long-range confinement effects surviving at finite $T$ in the non-Abelian plasma. The results reflect common features of Gribov-like approaches~\cite{Gribov:1977wm,Zwanziger:1989mf,Dudal:2008sp}, though the calculation was done via the GZ action.

\section{Conclusions and Outlook}\label{sec:c&o}

In this brief review, I have attempted to discuss recent progress of hard-thermal-loop effective theory on both the electric and magnetic scales of a hot QCD plasma. The HTLpt thermodynamics calculation has been a daunting task, and the NNLO thermodynamic potential at finite $T$ and $\mu$ resembles a continuous effort over the past 15 years. Much confidence has been gained from the results that HTLpt may provide a good description for QGP thermodynamic functions and various susceptibilities at $T\gtrsim300$\,MeV, and it would be interesting to apply the HTLpt framework to realtime quantities at these temperatures.

Comparing to the electric sector, the magnetic sector has been explored to much less degree. The uncovering of the massless Gribov mode has been an exciting attempt in exploring the significance of the magnetic scale to a non-Abelian plasma. It sheds new light on the active degrees of freedom released in course of a heavy-ion collision through which a strongly coupled QGP might emerge. It would be extremely tempting and challenging to explore the phenomenological significance of this new mode in interpreting experimental data.

\section*{Acknowledgments}

The author acknowledges Jens O. Andersen, Aritra Bandyopadhyay, Najmul Haque, Munshi G. Mustafa, Michael Strickland, and Konrad Tywoniuk for collaborations on which this review is based.


\begin{thebibliography}{0}

\bibitem{Gross:1973id}
  D.~J.~Gross and F.~Wilczek,
  {\it Phys. Rev. Lett.} {\bf 30}, 1343 (1973).

\bibitem{Politzer:1973fx}
  H.~D.~Politzer,
  {\it Phys. Rev. Lett.} {\bf 30}, 1346 (1973).
  
\bibitem{Schafer:2009dj}
  T.~Sch\"afer and D.~Teaney,
  {\it Rept. Prog. Phys.} {\bf 72}, 126001 (2009).
  
\bibitem{CasalderreySolana:2011us}
  J.~Casalderrey-Solana, H.~Liu, D.~Mateos, K.~Rajagopal and U.~A.~Wiedemann,
  arXiv:1101.0618 [hep-th].

\bibitem{Philipsen:2012nu}
  O.~Philipsen,
  {\it Prog. Part. Nucl. Phys.} {\bf 70}, 55 (2013).
  
\bibitem{Meyer:2011gj}
  H.~B.~Meyer,
  {\it Eur. Phys. J. A} {\bf 47}, 86 (2011).

\bibitem{Frenkel:1989br}
  J.~Frenkel and J.~C.~Taylor,
  {\it Nucl. Phys. B} {\bf 334}, 199 (1990).
  
\bibitem{Braaten:1989mz}
  E.~Braaten and R.~D.~Pisarski,
  {\it Nucl. Phys. B} {\bf 337}, 569 (1990).
  
\bibitem{Braaten:1990az}
  E.~Braaten and R.~D.~Pisarski,
  {Nucl. Phys. B} {\bf 339}, 310 (1990).
  
\bibitem{Taylor:1990ia}
  J.~C.~Taylor and S.~M.~H.~Wong,
  {\it Nucl. Phys. B} {\bf 346}, 115 (1990).
  
\bibitem{Frenkel:1991ts}
  J.~Frenkel and J.~C.~Taylor,
  {\it Nucl. Phys. B} {\bf 374}, 156 (1992).
  
\bibitem{Braaten:1991gm}
  E.~Braaten and R.~D.~Pisarski,
  {\it Phys. Rev. D} {\bf 45}, 1827 (1992).

\bibitem{Blaizot:2001nr}
  J.-P.~Blaizot and E.~Iancu,
  {\it Phys. Rept.} {\bf 359}, 355 (2002).
  
\bibitem{Blaizot:2003tw}
  J.-P.~Blaizot, E.~Iancu and A.~Rebhan,
  in {\it Quark Gluon Plasma}, edited by R. C. Hwa and X.-N. Wang (World Scientific, Singapore, 2004) p. 60.

\bibitem{Kraemmer:2003gd}
  U.~Kraemmer and A.~Rebhan,
  {\it Rept. Prog. Phys.}  {\bf 67}, 351 (2004).

\bibitem{Andersen:2004fp}
  J.~O.~Andersen and M.~Strickland,
  {\it Ann. Phys. (N.Y.)} {\bf 317}, 281 (2005).

\bibitem{Mrowczynski:2005ki}
  S.~Mr\'{o}wczy\'{n}ski,
  {\it Acta Phys. Polon. B} {\bf 37}, 427 (2006).

\bibitem{Arnold:2007pg}
  P.~B.~Arnold,
  {\it Int. J. Mod. Phys. E} {\bf 16}, 2555 (2007).

\bibitem{Su:2012iy}
  N.~Su,
  {\it Commun. Theor. Phys.} {\bf 57}, 409 (2012).

\bibitem{Linde:1980ts}
  A.~D.~Linde,
  {\it Phys. Lett. B} {\bf 96}, 289 (1980).

\bibitem{Gross:1980br}
  D.~J.~Gross, R.~D.~Pisarski and L.~G.~Yaffe,
  {\it Rev. Mod. Phys.} {\bf 53}, 43 (1981).
  
\bibitem{Shuryak:1977ut}
  E.~V.~Shuryak,
  {\it Sov. Phys. JETP} {\bf 47}, 212 (1978)
   [{\it Zh. Eksp. Teor. Fiz.} {\bf 74}, 408 (1978)].

\bibitem{Chin:1978gj}
  S.~A.~Chin,
  {\it Phys. Lett. B} {\bf 78}, 552 (1978).

\bibitem{Kapusta:1979fh}
  J.~I.~Kapusta,
  {\it Nucl. Phys. B} {\bf 148}, 461 (1979).

\bibitem{Toimela:1982hv}
  T.~Toimela,
  {\it Phys. Lett. B} {\bf 124}, 407 (1983).
  
\bibitem{Arnold:1994ps}
  P.~B.~Arnold and C.-X.~Zhai,
  {\it Phys. Rev. D} {\bf 50}, 7603 (1994).
  
\bibitem{Arnold:1994eb}
  P.~B.~Arnold and C.-X.~Zhai,
  {\it Phys. Rev. D} {\bf 51}, 1906 (1995).

\bibitem{Zhai:1995ac}
  C.-X.~Zhai and B.~M.~Kastening,
  {\it Phys. Rev. D} {\bf 52}, 7232 (1995).
  
\bibitem{Braaten:1995ju}
  E.~Braaten and A.~Nieto,
  {\it Phys. Rev. Lett.} {\bf 76}, 1417 (1996).
  
\bibitem{Braaten:1995jr}
  E.~Braaten and A.~Nieto,
  {\it Phys. Rev. D} {\bf 53}, 3421 (1996).
  
\bibitem{Kajantie:2002wa}
  K.~Kajantie, M.~Laine, K.~Rummukainen and Y.~Schr\"{o}der,
  {\it Phys. Rev. D} {\bf 67}, 105008 (2003).
  
 \bibitem{Andersen:1999fw}
  J.~O.~Andersen, E.~Braaten and M.~Strickland,
  {\it Phys. Rev. Lett.} {\bf 83}, 2139 (1999).
  
\bibitem{Karsch:1997gj}
  F.~Karsch, A.~Patk\'{o}s and P.~Petreczky,
  {\it Phys. Lett. B} {\bf 401}, 69 (1997).
  
\bibitem{Chiku:1998kd}
  S.~Chiku and T.~Hatsuda,
  {\it Phys. Rev. D} {\bf 58}, 076001 (1998).
  
\bibitem{Andersen:2000yj}
  J.~O.~Andersen, E.~Braaten and M.~Strickland,
  {\it Phys. Rev. D} {\bf 63}, 105008 (2001).

\bibitem{Andersen:2001ez}
  J.~O.~Andersen and M.~Strickland,
  {\it Phys. Rev. D} {\bf 64}, 105012 (2001).
  
\bibitem{Andersen:2008bz}
  J.~O.~Andersen and L.~Kyllingstad,
  {\it Phys. Rev. D} {\bf 78}, 076008 (2008).
  
\bibitem{Yukalov:1976pm}
  V.~I.~Yukalov,
  {\it Teor. Mat. Fiz.} {\bf 26}, 403 (1976).

\bibitem{Stevenson:1981vj}
  P.~M.~Stevenson,
  {\it Phys. Rev. D} {\bf 23}, 2916 (1981).

\bibitem{Duncan:1988hw}
  A.~Duncan and M.~Moshe,
  {\it Phys. Lett. B} {\bf 215}, 352 (1988).
  
\bibitem{Duncan:1992ba}
  A.~Duncan and H.~F.~Jones,
  {\it Phys. Rev. D} {\bf 47}, 2560 (1993).
  
\bibitem{Sisakian:1994nn}
  A.~N.~Sisakian, I.~L.~Solovtsov and O.~Shevchenko,
  {\it Int. J. Mod. Phys. A} {\bf 9}, 1929 (1994).

\bibitem{Janke:1995zz}
  W.~Janke and H.~Kleinert,
  {\it Phys. Rev. Lett.}  {\bf 75}, 2787 (1995).
  
\bibitem{Andersen:1999sf}
  J.~O.~Andersen, E.~Braaten and M.~Strickland,
  {\it Phys. Rev. D} {\bf 61}, 014017 (2000).
  
\bibitem{Andersen:1999va}
  J.~O.~Andersen, E.~Braaten and M.~Strickland,
  {\it Phys. Rev. D} {\bf 61}, 074016 (2000).
  
\bibitem{Andersen:2002ey}
  J.~O.~Andersen, E.~Braaten, E.~Petitgirard and M.~Strickland,
  {\it Phys. Rev. D} {\bf 66}, 085016 (2002).
  
\bibitem{Andersen:2003zk}
  J.~O.~Andersen, E.~Petitgirard and M.~Strickland,
  {\it Phys. Rev. D} {\bf 70}, 045001 (2004).
  
\bibitem{Andersen:2009tw}
  J.~O.~Andersen, M.~Strickland and N.~Su,
  {\it Phys. Rev. D} {\bf 80}, 085015 (2009).
  
\bibitem{Andersen:2009tc}
  J.~O.~Andersen, M.~Strickland and N.~Su,
  {\it Phys. Rev. Lett.} {\bf 104}, 122003 (2010).
  
\bibitem{Andersen:2010ct}
  J.~O.~Andersen, M.~Strickland and N.~Su,
  {\it JHEP} {\bf 1008}, 113 (2010).
  
\bibitem{Andersen:2010wu}
  J.~O.~Andersen, L.~E.~Leganger, M.~Strickland and N.~Su,
  {\it Phys. Lett. B} {\bf 696}, 468 (2011).
  
\bibitem{Andersen:2011sf}
  J.~O.~Andersen, L.~E.~Leganger, M.~Strickland and N.~Su,
  {\it JHEP} {\bf 1108}, 053 (2011).
  
\bibitem{Andersen:2011ug}
  J.~O.~Andersen, L.~E.~Leganger, M.~Strickland and N.~Su,
  {\it Phys. Rev. D} {\bf 84}, 087703 (2011).

\bibitem{Su:2011zv}
  N.~Su,
  arXiv:1104.3450 [hep-ph].
  
\bibitem{Andersen:2012wr}
  J.~O.~Andersen, S.~Mogliacci, N.~Su and A.~Vuorinen,
  {\it Phys. Rev. D} {\bf 87}, 074003 (2013).
  
\bibitem{Haque:2012my}
  N.~Haque, M.~G.~Mustafa and M.~Strickland,
  {\it Phys. Rev. D} {\bf 87}, 105007 (2013).
  
\bibitem{Haque:2013qta}
  N.~Haque, M.~G.~Mustafa and M.~Strickland,
  {\it JHEP} {\bf 1307}, 184 (2013).
  
\bibitem{Mogliacci:2013mca}
  S.~Mogliacci, J.~O.~Andersen, M.~Strickland, N.~Su and A.~Vuorinen,
  {\it JHEP} {\bf 1312}, 055 (2013).
  
\bibitem{Haque:2013sja}
  N.~Haque, J.~O.~Andersen, M.~G.~Mustafa, M.~Strickland and N.~Su,
  {\it Phys. Rev. D} {\bf 89}, 061701 (2014).
  
\bibitem{Haque:2014rua}
  N.~Haque, A.~Bandyopadhyay, J.~O.~Andersen, M.~G.~Mustafa, M.~Strickland and N.~Su,
  {\it JHEP} {\bf 1405}, 027 (2014).
  
\bibitem{Baier:1999db}
  R.~Baier and K.~Redlich,
  {\it Phys. Rev. Lett.} {\bf 84}, 2100 (2000).

\bibitem{Andersen:2002jz}
  J.~O.~Andersen and M.~Strickland,
  {\it Phys. Rev. D} {\bf 66}, 105001 (2002).
  
\bibitem{Chakraborty:2001kx}
  P.~Chakraborty, M.~G.~Mustafa and M.~H.~Thoma,
  {\it Eur. Phys. J. C} {\bf 23}, 591 (2002).
  
\bibitem{Chakraborty:2002yt}
  P.~Chakraborty, M.~G.~Mustafa and M.~H.~Thoma,
  {\it Phys. Rev. D} {\bf 67}, 114004 (2003).
  
\bibitem{Chakraborty:2003uw}
  P.~Chakraborty, M.~G.~Mustafa and M.~H.~Thoma,
  {\it Phys. Rev. D} {\bf 68}, 085012 (2003).
  
\bibitem{Haque:2011iz}
  N.~Haque, M.~G.~Mustafa and M.~H.~Thoma,
  {\it Phys. Rev. D} {\bf 84}, 054009 (2011).

\bibitem{Haque:2010rb}
  N.~Haque and M.~G.~Mustafa,
  arXiv:1007.2076 [hep-ph].

\bibitem{Blaizot:1999ip}
  J.-P.~Blaizot, E.~Iancu and A.~Rebhan,
  {\it Phys. Rev. Lett.} {\bf 83}, 2906 (1999).
  
\bibitem{Blaizot:1999ap}
  J.-P.~Blaizot, E.~Iancu and A.~Rebhan,
  {\it Phys. Lett. B} {\bf 470}, 181 (1999).
  
\bibitem{Blaizot:2000fc}
  J.-P.~Blaizot, E.~Iancu and A.~Rebhan,
  {\it Phys. Rev. D} {\bf 63}, 065003 (2001).

\bibitem{Blaizot:2001vr}
  J.-P.~Blaizot, E.~Iancu and A.~Rebhan,
  {\it Phys. Lett. B} {\bf 523}, 143 (2001).

\bibitem{Blaizot:2002xz}
  J.-P.~Blaizot, E.~Iancu and A.~Rebhan,
  {\it Eur. Phys. J. C} {\bf 27}, 433 (2003).
  
\bibitem{Vuorinen:2003fs}
  A.~Vuorinen,
  {\it Phys. Rev. D} {\bf 68}, 054017 (2003).
  
\bibitem{Bazavov:2012ka}
  A.~Bazavov {\it et al.}, 
  {\it Phys. Rev. D} {\bf 86}, 114031 (2012).
  
\bibitem{Bazavov:2009zn}
  A.~Bazavov {\it et al.}, 
  {\it Phys. Rev. D} {\bf 80}, 014504 (2009).
  
\bibitem{Borsanyi:2010cj}
  S.~Bors\'anyi {\it et al.}, 
  {\it JHEP} {\bf 1011}, 077 (2010).
  
\bibitem{Borsanyi:2012cr}
  S.~Bors\'anyi {\it et al.}, 
  {\it JHEP} {\bf 1208}, 053 (2012).
  
\bibitem{Bazavov:2013dta}
  A.~Bazavov {\it et al.}, 
  {\it Phys. Rev. Lett.} {\bf 111}, 082301 (2013).
  
\bibitem{Bazavov:2013uja}
  A.~Bazavov {\it et al.}, 
  {\it Phys. Rev. D} {\bf 88}, 094021 (2013).
  
\bibitem{Bazavov:2012vg}
  A.~Bazavov {\it et al.}, 
  {\it Phys. Rev. Lett.} {\bf 109}, 192302 (2012).
  
\bibitem{Datta:2014zqa}
  S.~Datta, R.~V.~Gavai and S.~Gupta,
  {\it PoS LATTICE} {\bf 2013}, 202 (2014).
  
\bibitem{Borsanyi:2011sw}
  S.~Bors\'anyi {\it et al.}, 
  {\it JHEP} {\bf 1201}, 138 (2012).
  
\bibitem{Bernard:2004je}
  C.~Bernard {\it et al.}  [MILC Collaboration],
  {\it Phys. Rev. D} {\bf 71}, 034504 (2005).
  
\bibitem{Buchmuller:1994qy}
  W.~Buchm\"uller and O.~Philipsen,
  {\it Nucl. Phys. B} {\bf 443}, 47 (1995).
  
\bibitem{Alexanian:1995rp}
  G.~Alexanian and V.~P.~Nair,
  {\it Phys. Lett. B} {\bf 352}, 435 (1995).
  
\bibitem{Jackiw:1995nf}
  R.~Jackiw and S.~Y.~Pi,
  {\it Phys. Lett. B} {\bf 368}, 131 (1996).
  
\bibitem{Jackiw:1997jga}
  R.~Jackiw and S.~Y.~Pi,
  {\it Phys. Lett. B} {\bf 403}, 297 (1997).
  
\bibitem{Cornwall:1997dc}
  J.~M.~Cornwall,
  {\it Phys. Rev. D} {\bf 57}, 3694 (1998).
  
\bibitem{Eberlein:1998yk}
  F.~Eberlein,
  {\it Phys. Lett. B} {\bf 439}, 130 (1998).
  
\bibitem{Bieletzki:2012rd}
  D.~Bieletzki, K.~Lessmeier, O.~Philipsen and Y.~Schr\"oder,
  {\it JHEP} {\bf 1205}, 058 (2012).
  
\bibitem{Alkofer:2000wg}
  R.~Alkofer and L.~von Smekal,
  {\it Phys. Rept.}  {\bf 353}, 281 (2001).
  
\bibitem{Cucchieri:2004mf}
  A.~Cucchieri, T.~Mendes and A.~R.~Taurines,
  {\it Phys. Rev. D} {\bf 71}, 051902 (2005).
  
\bibitem{Bowman:2007du}
  P.~O.~Bowman {\it et al.}, 
  {\it Phys. Rev. D} {\bf 76}, 094505 (2007).
  
\bibitem{Silva:2014sea}
  P.~J.~Silva, O.~Oliveira, D.~Dudal, P.~Bicudo and N.~Cardoso,
  {\it PoS} ({\bf QCD-TNT-III}), 040 (2013),
  arXiv:1401.1554 [hep-lat].
  
\bibitem{Alkofer:2003jj}
  R.~Alkofer, W.~Detmold, C.~S.~Fischer and P.~Maris,
  {\it Phys. Rev. D} {\bf 70}, 014014 (2004).
  
\bibitem{Maas:2004se}
  A.~Maas, J.~Wambach, B.~Gruter and R.~Alkofer,
  {\it Eur. Phys. J. C} {\bf 37}, 335 (2004).
  
\bibitem{Maas:2005hs}
  A.~Maas, J.~Wambach and R.~Alkofer,
  {\it Eur. Phys. J. C} {\bf 42}, 93 (2005).
  
\bibitem{Fischer:2008uz}
  C.~S.~Fischer, A.~Maas and J.~M.~Pawlowski,
  {\it Annals Phys.}  {\bf 324}, 2408 (2009).
  
\bibitem{Fister:2011uw}
  L.~Fister and J.~M.~Pawlowski,
  arXiv:1112.5440 [hep-ph].
  
\bibitem{Strauss:2012dg}
  S.~Strauss, C.~S.~Fischer and C.~Kellermann,
  {\it Phys. Rev. Lett.}  {\bf 109}, 252001 (2012).
  
\bibitem{Maas:2005ym}
  A.~Maas,
  {\it Mod. Phys. Lett. A} {\bf 20}, 1797 (2005).
  
\bibitem{Maas:2011se}
  A.~Maas,
  {\it Phys. Rept.}  {\bf 524}, 203 (2013).

\bibitem{Harada:2008vk}
  M.~Harada and Y.~Nemoto,
  {\it Phys. Rev. D} {\bf 78}, 014004 (2008).
  
\bibitem{Qin:2010pc}
  S.-X.~Qin, L.~Chang, Y.-X.~Liu and C.~D.~Roberts,
  {\it Phys. Rev. D} {\bf 84}, 014017 (2011).
  
\bibitem{Nakkagawa:2011ci}
  H.~Nakkagawa, H.~Yokota and K.~Yoshida,
  {\it Phys. Rev. D} {\bf 85}, 031902 (2012).
  
\bibitem{Nakkagawa:2012ip}
  H.~Nakkagawa, H.~Yokota and K.~Yoshida,
  {\it Phys. Rev. D} {\bf 86}, 096007 (2012).
  
\bibitem{Gao:2014rqa}
  F.~Gao, S.-X.~Qin, Y.-X.~Liu, C.~D.~Roberts and S.~M.~Schmidt,
  {\it Phys. Rev. D} {\bf 89}, 076009 (2014).

\bibitem{Kitazawa:2005mp}
  M.~Kitazawa, T.~Kunihiro and Y.~Nemoto,
  {\it Phys. Lett. B} {\bf 633}, 269 (2006).
  
\bibitem{Kitazawa:2006zi}
  M.~Kitazawa, T.~Kunihiro and Y.~Nemoto,
  {\it Prog. Theor. Phys.} {\bf 117}, 103 (2007).
  
\bibitem{Kitazawa:2007ep}
  M.~Kitazawa, T.~Kunihiro, K.~Mitsutani and Y.~Nemoto,
  {\it Phys. Rev. D} {\bf 77}, 045034 (2008).
  
\bibitem{Hidaka:2011rz}
  Y.~Hidaka, D.~Satow and T.~Kunihiro,
  {\it Nucl. Phys. A} {\bf 876}, 93 (2012).
  
\bibitem{Satow:2013oya}
  D.~Satow,
  {\it Phys. Rev. D} {\bf 87}, 096011 (2013).
  
\bibitem{Blaizot:2014hka}
  J.-P.~Blaizot and D.~Satow,
  {\it Phys. Rev. D} {\bf 89}, 096001 (2014).
  
\bibitem{Su:2014rma}
  N.~Su and K.~Tywoniuk,
  arXiv:1409.3203 [hep-ph].
  
\bibitem{Gribov:1977wm}
  V.~N.~Gribov,
  {\it Nucl. Phys. B} {\bf 139}, 1 (1978).
  
\bibitem{Zwanziger:1989mf}
  D.~Zwanziger,
  {\it Nucl. Phys. B} {\bf 323}, 513 (1989).
  
\bibitem{Dokshitzer:2004ie}
  Y.~L.~Dokshitzer and D.~E.~Kharzeev,
  {\it Ann. Rev. Nucl. Part. Sci.} {\bf 54}, 487 (2004).
  
\bibitem{Sobreiro:2005ec}
  R.~F.~Sobreiro and S.~P.~Sorella,
  hep-th/0504095.
  
\bibitem{Vandersickel:2012tz}
  N.~Vandersickel and D.~Zwanziger,
  {\it Phys. Rept.} {\bf 520}, 175 (2012).

\bibitem{Zwanziger:2006sc}
  D.~Zwanziger,
  {\it Phys. Rev. D} {\bf 76}, 125014 (2007).
  
\bibitem{Fukushima:2013xsa}
  K.~Fukushima and N.~Su,
  {\it Phys. Rev. D} {\bf 88}, 076008 (2013).
  
\bibitem{Zwanziger:2004np}
  D.~Zwanziger,
  {\it Phys. Rev. Lett.} {\bf 94}, 182301 (2005).
  
\bibitem{Lichtenegger:2008mh}
  K.~Lichtenegger and D.~Zwanziger,
  {\it Phys. Rev. D} {\bf 78}, 034038 (2008).
  
\bibitem{Kojo:2009ha}
  T.~Kojo, Y.~Hidaka, L.~McLerran and R.~D.~Pisarski,
  {\it Nucl. Phys. A} {\bf 843}, 37 (2010).
  
\bibitem{Kojo:2011cn}
  T.~Kojo, Y.~Hidaka, K.~Fukushima, L.~D.~McLerran and R.~D.~Pisarski,
  {\it Nucl. Phys. A} {\bf 875}, 94 (2012).
  
\bibitem{Kojo:2012js}
  T.~Kojo and N.~Su,
  {\it Phys. Lett. B} {\bf 720}, 192 (2013).
  
\bibitem{Kaczmarek:2005ui}
  O.~Kaczmarek and F.~Zantow,
  {\it Phys. Rev. D} {\bf 71}, 114510 (2005).
  
\bibitem{Dudal:2008sp}
  D.~Dudal, J.~A.~Gracey, S.~P.~Sorella, N.~Vandersickel and H.~Verschelde,
  {\it Phys. Rev. D} {\bf 78}, 065047 (2008).

\end{thebibliography}
\end{document}